\documentclass[aps,pre,floatfix,superscriptaddress,twocolumn,10pt]{revtex4-1}
\usepackage{ulem}
\usepackage{color}
\usepackage{graphicx}
\usepackage{amsmath, amsthm, amssymb}
\usepackage[colorlinks,citecolor=red,urlcolor=blue]{hyperref}
\newcommand{\be}{\begin{equation}}
\newcommand{\ee}{\end{equation}}
\newcommand{\bea}{\begin{eqnarray}}
\newcommand{\eea}{\end{eqnarray}}
\newcommand{\Qn}{\mathbf{Q}}

\begin{document}
\title{Different phases of a system of hard rods on three dimensional cubic lattice}

\author{N. Vigneshwar}
\email{vigneshwarn@imsc.res.in}
\affiliation{The Institute of Mathematical Sciences, C.I.T. Campus, Taramani, Chennai 600113, India}
\affiliation{Homi Bhabha National Institute, Training School Complex, Anushakti Nagar, Mumbai 400094, India}
\author{Deepak Dhar} 
\email{deepakdhar1951@gmail.com}
\affiliation{Indian Institute of Science Education and Research, Dr. Homi Bhabha Road, Pashan,  Pune 411008, India}
\author{R. Rajesh}
\email{rrajesh@imsc.res.in}
\affiliation{The Institute of Mathematical Sciences, C.I.T. Campus, Taramani, Chennai 600113, India}
\affiliation{Homi Bhabha National Institute, Training School Complex, Anushakti Nagar, Mumbai 400094, India}

\date{\today}

\begin{abstract}
We study the different phases of a system of monodispersed hard rods of length $k$ on a cubic lattice using an efficient cluster algorithm which can simulate densities  close to the  fully-packed limit. For $k\leq 4$, the system is disordered at all densities. For $k=5,6$, we find a single density-driven transition from a disordered phase to high density layered-disordered phase in which the density of
rods of one orientation is strongly  suppressed, breaking the system into weakly coupled layers. Within a layer, the system is disordered. For $k \geq 7$,  three density driven transitions are observed numerically: isotropic to nematic to layered-nematic to layered-disordered.   In the layered-nematic phase, the system 
breaks up into layers, with nematic order in each  in each layer, but very weak correlation between the ordering direction between different layers. We argue that the layered-nematic phase is a finite-size effect, and  in the thermodynamic limit, the nematic phase will have higher entropy per site.

\end{abstract}

\maketitle
\section{Introduction}

The study of entropy driven phase transitions in systems of long hard rods has a long history dating back to Onsager's demonstration
of a phase transition from an isotropic phase to an orientationally ordered nematic phase with increasing density~\cite{1949-o-nyas-effects}.
At even higher densities, a system of spherocylinders will show a smectic/columnar phases with partial translational order along
with orientational order~\cite{1997-pf-pre-first, 1988-fls-nature-thermodynamic} and solid-like phases~\cite{1997-bf-jcp-tracing}. 
In two dimensions, there is no true ordered phase, but a system of needles undergoes
a Kosterlitz-Thouless type transition into a high density phase with power law correlations~\cite{2005-kb-pre-orientational,1985-fe-pra-evidence,2009-v-epjb-isotropic}. 
Physical systems where such transitions are observed include  aqueous solutions tobacco mosaic viruses~\cite{1989-fmcm-prl-isotropic}, 
liquid crystals~\cite{1995-dp-oup-physics}, oxygen monolayers adsorbed on molybdenum surfaces~\cite{1991-dmr-jcp-model}, carbon 
nanotube nematic gels~\cite{2004-iadzly-prl-nematic} and chlorine atoms adsorbed   on silver~\cite{2001-mbr-ss-static}.

The corresponding problem on lattices where the orientation of rods are restricted to the lattice directions 
(also known as Zwanzig model) has also been studied in parallel. Consider a  system  of monodispersed
rods of length $k$ that occupy $k$ consecutive lattice sites along any one of the lattice directions. Two rods cannot overlap.
Early work based on virial expansion~\cite{1963-z-jcp-first}, high density expansions~\cite{1956-f-rspa-statistical} and the Guggenheim approximation~\cite{1961-d-jcp-statistics}, predicted a transition from the low-density disordered phase to a nematic-ordered phase, as the density is increased.  
However, not much is known about the nature of the high-density phase in this model. At densities near full packing,  nematic order is not expected to survive, as there are exponentially many disordered configurations \cite{1995-dp-oup-physics,2007-gd-epl-on}.  In two dimensions, numerical simulations have shown that the high-density disordered phase has no orientational long-range order, but  it is not clear whether some  subtle kind of order exists or not.  In three dimensions, what is the nature of different phases in the lattice model of hard rods? This is the primary question that we address in this paper. We find that the high-density phase does not have nematic ordering, where rods preferentially align themselves parallel to each other. Instead,  for $k>4$, they develop a smectic-like order, where {\it two} of the three orientations have large, nearly equal values, and the density of rods of  third orientation is very small and
the different layers are nearly independent. This way, entropy is maximised,
while satisfying the packing constraints at high density. We do obtain a nematic phase, but only at intermediate densities, and only for $k\geq 7$. The determination of the rich structure of different possible orderings in this simple model is the main new finding of this paper.  We expect that the behavior in dimensions $d > 3$ would be qualitatively similar. 

We first summarize known results for the model in two and three dimensions.
In two dimensions, for $k=2$ (dimers), it may be shown rigorously that the system is disordered at all densities~\cite{1972-hl-cmp-theory}, while at full packing, the system is power-law correlated ~\cite{1963-fs-pr-statistical,2003-hkms-prl-coulomb}. For $k\leq 6$, Monte Carlo
simulations show that the system is disordered at all densities~\cite{2007-gd-epl-on}. For $k \geq 7$, the system undergoes two
transitions: first from a low-density disordered phase to an intermediate density nematic phase, and second from the nematic
phase to a high density disordered phase~\cite{2007-gd-epl-on,2013-krds-pre-nematic,2013-kr-pre-reentrant}. While the first transition belongs
to the Ising and 3-state Potts universality classes on square~\cite{2008-mlr-epl-determination, 2008-mlr-jcp-critical,2009-fv-epl-restricted} and triangular lattices~\cite{2008-mlr-jcp-critical,2008-mlr-pa-critical} respectively, the universality class of the
second transition has been difficult to resolve~\cite{2013-krds-pre-nematic,2013-kr-pre-reentrant}. For large enough $k$, the
existence of the nematic phase may be rigorously proved~\cite{2013-dg-cmp-nematic}. The model may also be solved exactly on a tree-like
lattice, corresponding to a Bethe approximation, and shows a nematic phase for $k \geq 4$, but
does not exhibit a second transition~\cite{2011-drs-pre-hard}. Density functional theory for rods give a similar result~\cite{2016-okdesh-jcp-monolayers}.
The problem of rods may also be generalized to
hard rectangles of size $m \times d$ or hard squares of size $m \times m$. For large aspect ratio, the system of rectangles
shows four phases with increasing densities: disordered to nematic to columnar to solid-like phases. The detailed phase
diagrams for different choices of $m$ and $d$, and the asymptotic behavior of the phase boundaries may be found in Ref.~\cite{2014-kr-pre-phase,2015-kr-epjb-phase,2015-kr-pre-asymptotic,2015-nkr-jsp-high}. The $m \times m$ hard square 
models undergo a single density-driven transition from a disordered phase to a high-density columnar 
phase~\cite{1967-bn-jcp-phase,1966-rc-jcp-phase,2007-fal-jcp-monte,2007-zt-prb-lattice,2011-fbn-pre-lattice,
2012-rd-pre-high,2015-rdd-prl-columnar,2016-ndr-epl-stability,2017-mnr-jsm-estimating,2017-mr-arxiv-columnar}.

In three dimensions, 
dimer models ($k=2$) at full packing  on bipartite lattices are known to show a Coulomb phase, with algebraic decay of orientational correlations ~\cite{2003-hkms-prl-coulomb}, while for non-bipartite lattices, the correlations decay faster, and in some exactly solved cases,  correlations are strictly zero beyond a finite range ~\cite{2008-dc-prl-exact}.   Not much is known for larger values of $k$. It would be expected that, like
in the continuum, there will be an isotropic-nematic transition as the density is increased above zero.  However,
the minimum value of $k$ is for such a transition to occur is not known. Also, the nature of ordering at still higher densities  is not studied much.  

In this paper, we study the problem of a monodispersed system of rods of length $k$ using grand canonical Monte Carlo 
simulations that implements an algorithm with cluster moves. For $k\leq4$, we find that the system remains disordered and is in the
isotropic phase at all densities. When $k=5, 6$, we observe a single transition from a low density disordered phase 
in which the the fractional number of different orientations is nearly equal,  to a layered-disordered phase in which the fractional number of one orientation becomes very small, and system develops a layer-like structure, where each layer is a plane with most of the rods being of two orientations lying within the plane, and very weak correlations between different layers. When $k\geq7$, at intermediate densities, we numerically observe two other phases:  a nematic phase, and  a new phase that we call  the layered-nematic phase. In the  layered-nematic phase, each plane 
has  two dimensional nematic order, but there is no overall bulk nematic order.  We 
argue that the existence of this layered-nematic phase is a finite size effect, and that in the thermodynamic limit, parallel orientation between different layers will be regained.

The rest of the paper is organized as follows. In Sec.~\ref{sec:model}, we define the model precisely and describe the 
grand canonical Monte Carlo scheme that is used to simulate the system. Section~\ref{sec:phases} describes the different
phases -- isotropic, nematic, layered-nematic and layered-disordered -- that we observe in our simulations. In Sec.~\ref{sec:stability}, we  use perturbation theory to argue that the layered-nematic phase observed in simulations is an artifact of finite system sizes, and the observed behavior should cross over to nematic order for length-scales  greater that some crossover scale $L^*( \rho)$, where $\rho$ is the density of covered sites. 
Section~\ref{sec:phasediagram} consists results of detailed simulations for systems with $k=2,3,\ldots,7$. The
minimum length of rods that is needed for each of the phases to exist is determined. The critical densities and chemical
potentials, and other critical parameters are determined for $k=5,6,7$. We end with a summary and discussion of results in Sec.~\ref{sec:discussion}.

\section{\label{sec:model}Model Description and Monte-Carlo Algorithm}
 
Consider a cubic lattice of size $L\times L\times L$ with periodic boundary conditions. The lattice sites may be occupied by 
rods  that occupy $k$ consecutive lattices sites in any one of the three mutually orthogonal directions. 
The rods interact only through excluded volume interactions, i.e., a lattice site may be occupied by at most one rod.  
We associate a weight $e^\mu$ with each rod, where $\mu$ is the chemical potential rescaled by temperature. We will call
a rod oriented 
in the $x$-, $y$- and $z$-directions as $x$-mer, $y$-mer, and $z$-mer respectively. 
The site of a rod with the smallest $x$-, $y$-, and $z$-coordinates
will be called its head.

The aim of the paper is to determine the different phases in the hard rod model as the density, or the rod-length $k$ is varied.  
This is done primarily through grand canonical Monte Carlo simulations. 
Conventional algorithms with local evaporation and deposition moves fail to equilibrate the system 
(within available computer time) at large densities because the system gets stuck in long-lived metastable states. 
Instead, we implement a Monte Carlo algorithm with cluster moves~\cite{2012-krds-aipcp-monte,2013-krds-pre-nematic} that has recently proved useful in equilibrating systems of hard particles with large excluded volume interactions at densities 
close to one~\cite{2012-krds-aipcp-monte,2013-krds-pre-nematic} or at full packing~\cite{2015-rdd-prl-columnar}. 

The Monte Carlo algorithm that we use is the following: remove all the $x$-mers, leaving all $y$-mers and $z$-mers
undisturbed. The empty intervals in each row in the $x$-direction, separated from each other by $y$-mers or 
$z$-mers, is now re-occupied by $x$-mers with the correct equilibrium probabilities. The calculation of these probabilities
reduces to a one dimensional problem which may be solved exactly (see Refs.~\cite{2013-krds-pre-nematic, 2015-rdd-prl-columnar, 2014-nr-pre-multiple}  for details). The evaporation and deposition move  satisfies detailed balance  as the transition rates depend only on the equilibrium probabilities of the new configuration. Following evaporation and deposition of $x$-mers, we repeat the set of steps  with $y$-mers, and then with $z$-mers.

\begin{figure}
\includegraphics[width=0.6\columnwidth]{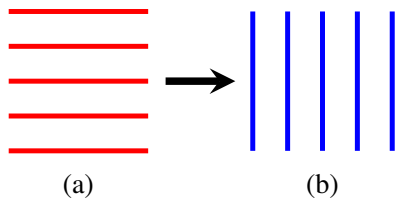}
\caption{\label{flipmove}A schematic diagram illustrating the flip move in the Monte Carlo algorithm. If there is a $k \times k$ square, that is fully covered by 
$k$ parallel $k$-mers as shown in (a), then the orientations of the $k$-mers within the square are flipped to the configuration  shown in (b).  }
\end{figure}

To reduce equilibration and autocorrelation times at high densities, we also implement a flip move.
If there is a $k \times k$ square, that is fully covered by $k$ parallel $k$-mers, then we can flip the orientation of $k$-mers, within this square, without affecting any other rods, as shown in the schematic diagram in Fig.~\ref{flipmove}.
Clearly, the flip move  does not violate the hard-core constraint and satisfies detailed balance.
We define one Monte Carlo time step as updating every 
row in the $x$-, $y$- and $z$- directions (total of $3 L^2$ rows),
and $L^3$(in case of small system sizes) or $L^3/k^2$(in case of large system sizes) flip moves.

\begin{figure}
\includegraphics[width=\columnwidth]{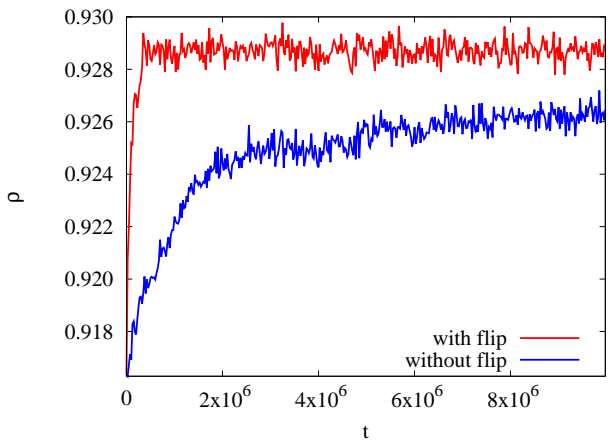}
\caption{\label{comp}The temporal evolution of the density $\rho$ in a system with $k=7$, when the system is evolved using the evaporation-deposition algorithm with and without the flip move. The initial configuration has nematic order while the equilibrium configuration has layered order. 
The data are for system size  $L=112$ and $\mu=6.0$. }
\end{figure}

The flip move is crucial for equilibrating the system at densities close to full packing. Figure~\ref{comp} shows the time
evolution of density $\rho$ for a system with $k=7$, starting from nematic initial conditions in which most of the rods lie in the $x$-direction,
using the evaporation-deposition algorithm with and without the flip move. The value of $\mu$ is such that the equilibrium configuration does not have nematic order (see Sec.~\ref{sec:phasediagram} for details). When the flip move is present, $\rho$ reaches its
equilibrium value in about  $3\times 10^5$ Monte Carlo steps. On the other hand, when the flip move is absent, 
the system does not reach equilibrium  
even after $10^7$ Monte Carlo steps. 

The algorithm
is easily parallelized as all the  rows can be updated simultaneously. The flip move may also be parallelized by 
choosing a plane and then choosing one of the $k^2$ sublattices randomly. All $k \times k$ squares with left bottom corner lying in this
sublattice may be updated simultaneously. All the data presented in the paper is obtained through a parallelized implementation of
the algorithm.

\section{\label{sec:phases}Different phases}

In this section, we  describe and define the different phases that we observe in our Monte Carlo simulations. Let $\rho_x$, $\rho_y$ and $\rho_z$ be the density of sites occupied by $x$-mers, $y$-mers and $z$-mers respectively. Consider the two-dimensional vector
\begin{equation}
   \Qn = |\Qn|e^{i\theta} = \rho_x + \rho_y\,e^{\frac{2\pi i}{3}} + \rho_z\,e^{\frac{4\pi i}{3}}.
\end{equation}
We define the nematic order parameters as
\begin{eqnarray}
\label{eq:qn}
  Q_N &=& \langle|\Qn|\rangle,\\
  P_2 &=& \langle\cos{(3\theta)}\rangle,
  \label{eq:p2}
\end{eqnarray}
where $\langle\cdots\rangle$ denotes average over the equilibrium probabilities and $\rho = \rho_x+\rho_y+\rho_z$ is the total fraction of sites occupied by $k$-mers.

\textit{Isotropic phase:} In the isotropic phase, the system is disordered with $\rho_x\approx \rho_y\approx \rho_z$. The probability distribution of $\Qn$ is centered about the origin and the order parameters take the value $Q_N\approx0$ and $P_2\approx0$.

\begin{figure}
\includegraphics[width=\columnwidth]{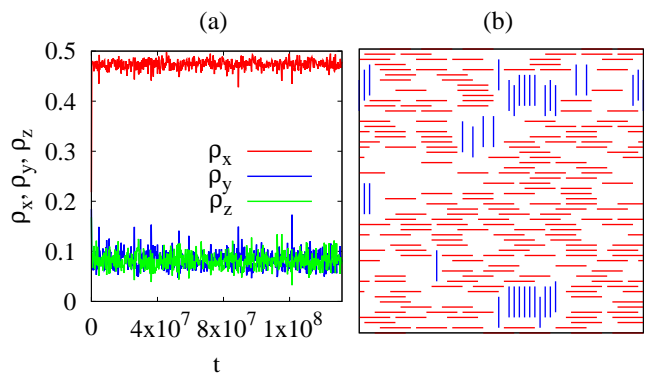}
\caption{\label{nem_1}(a) Time evolution of the densities of rods along the three orientations in the nematic phase when $\mu=0.3$ for $k=7$ and $L=56$. The initial configuration
is disordered. (b) Snapshot of a randomly chosen $xy$ plane after equilibration, showing dominance of $x$-mers.}
\end{figure}

\textit{Nematic phase:} In the nematic phase, a majority of the rods are of one orientation, while the rods of the other two orientations have smaller, roughly equal densities. If $x$- is the preferred direction, then $\rho_x \gg \rho_y \approx \rho_z$, as can be seen in the temporal
evolution of three densities shown in  Fig.~\ref{nem_1}(a). A snapshot of a randomly chosen $xy$ plane, as shown in Fig.~\ref{nem_1}(b), clearly shows that most rods are $x$-mers. In the nematic phase, $Q_N \approx \rho$ and $P_2\approx1$.

\begin{figure}
\includegraphics[width=\columnwidth]{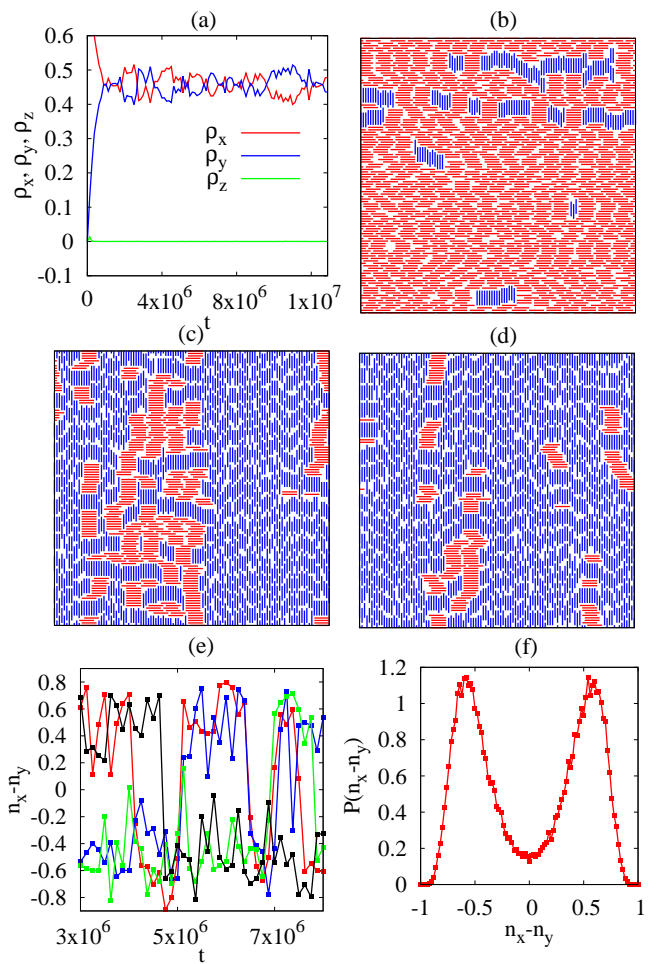}
\caption{\label{ln_1}(a) Time evolution of the densities of rods along the three orientations in the layered-nematic phase 
when $\mu=5.55$ for $k=7$ and $L=112$.  The initial configuration has nematic order, where most of the rods are in $x$-direction. 
(b)--(d) Snapshots of three randomly chosen $xy$ planes after equilibration. In each of the planes, either horizontal or vertical rods are in majority. 
(e) Time evolution of  $n_x(z)-n_y(z)$, where $n_x(z)$ and $n_y(z)$ are the densities of $x$-mers and $y$-mers
is layer $z$, for $z=0,24,49,74$. The nematic order in each plane keeps switching between majority $x$-mers and majority $y$-mers.  
(d) The probability distribution $ P( n_x(z)-n_y(z))$,
averaged over time and all planes, exhibits two symmetric peaks. } 
\end{figure}

\textit{Layered-Nematic phase:} In the layered-nematic phase, there is a spontaneous symmetry breaking, and one of the $xy$, $yz$ or $zx$ planes 
is selected, and the density of rods that are oriented perpendicular to this plane is suppressed [see Fig.~\ref{ln_1}(a)], making the system
layered. If the chosen plane is the $xy$ plane,
then $\rho_x\approx\rho_y\gg\rho_z$. In the layered-nematic phase, within a $xy$-plane, 
the rods have two-dimensional nematic order. This may  seen in the snapshots, shown in Fig.~\ref{ln_1}(b)--(d), of three randomly chosen $xy$ planes. 
Each of the planes has two-dimensional nematic order, but could be majority $x$-mers or$y$-mers. 
To quantify further, we show the time evolution of the local nematic order parameter $n_x(z)-n_y(z)$, where $n_x(z)$ and $n_y(z)$ are the densities of sites occupied by $x$-mers and $y$-mers in layer $z$, for four planes  in Fig.~\ref{ln_1}(e). Each of the planes has nematic order
which could change sign during the time evolution. 
There are roughly equal number of planes with majority $x$-mers and majority $y$-mers, as
may be seen from the double-peaked probability distribution function $P(n_x-n_y)$, shown in Fig.~\ref{ln_1}(f), 
which is obtained by averaging
over the different $xy$ planes  and over time. If the system has layered-nematic phase, $Q_N\approx \rho/2$ and $P_2\approx-1$.

\begin{figure}
\includegraphics[width=\columnwidth]{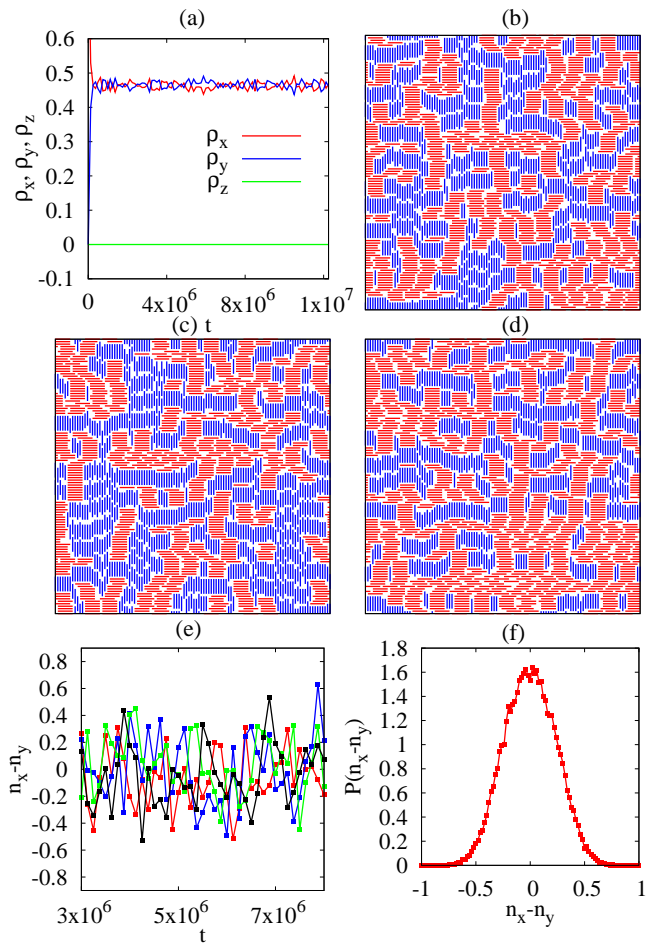}
\caption{\label{ld_1} (a) Time evolution of the densities of rods along the three orientations in the layered-disordered phase when $\mu=6.0$ for $k=7$ and $L=112$. The initial configuration has nematic order, where most of the rods are in $x$-direction. (b)--(d) Snapshots of three randomly chosen $xy$ planes after equilibration. In each of the planes, there are roughly equal number of $x$-mers and $y$-mers. (e) Time evolution of  $n_x(z)-n_y(z)$, where $n_x(z)$ and $n_y(z)$ are the densities of $x$-mers and $y$-mers
is layer $z$, for $z=0,24,49,74$. It fluctuates about zero for all $z$. d) The probability distribution $ P( n_x(z)-n_y(z))$,
averaged over time and all planes, is peaked about $0$. }
\end{figure}

\textit{Layered-Disordered phase:} In the layered-disordered phase, like in the layered-nematic phase, majority of the rods 
lie in one of the $xy$, $yz$ or $zx$ planes [see Fig.~\ref{ld_1}(a)]. Let the chosen plane be the $xy$ plane, i.e.,  $\rho_x\approx\rho_y\gg\rho_z$. In the layered-disordered phase, unlike the layered-nematic phase, the rods within a $xy$ plane do not have nematic order, i.e., $n_x(z)\approx n_y(z)$ for each layer $z$. This may  seen in the snapshots, shown in Fig.~\ref{ld_1}(b)--(d), of three randomly chosen $xy$ planes, where in each of the planes, there are  roughly equal number of $x$-mers  and $y$-mers present. 
The nematic order in each plane fluctuates about zero, as may be seen from the time evolution of the nematic order of four
planes as shown in  Fig.~\ref{ld_1}(e)] as well as probability distribution [see Fig.~\ref{ld_1}(f)] of the
local nematic order parameter $n_x(z)-n_y(z)$. 
In the layered-disordered phase, $Q_N\approx \rho/2$ and $P_2\approx-1$. 

The order parameters $Q_N$ and $P_2$ fail to distinguish between the layered-nematic and layered-disordered phases and take the values $Q_N\approx \rho/2$ and $P_2\approx-1$ for both phases. Though we observe both these phases in our simulations, we argue in the next section that the layered-nematic phase has lower entropy per site that the nematic phase, and is thus a metastable phase.

\section{\label{sec:stability}The instability of the layered-nematic phase}

In this section, we discuss the instability of the layered-nematic phase. We argue that  the layered-nematic phase seen in our simulations is the result of finite size of our samples, and different layers  would be expected to develop alignment and hence the usual nematic order if we could study  samples of much larger  sizes.

We start by considering a system in which the chemical potential of rods  in different directions are different: in the $x$- and $y$-directions, it is $\mu$, but in the $z$-direction it is $\mu'$.  We start with the case when $z' = e^{\mu'}=0$. In this case, 
different $z$-layers decouple, and the problem reduces to the problem of $k$-mers on a two dimensional square lattice. 
We assume that $\mu$ is such that in each layer there is nematic ordering, but in different layers, it may be in different directions.  We consider the spontaneous-symmetry broken state $\{ \sigma\}$, where the ordering direction in the layer $z=i$ is $\sigma_i$, taking values $\pm 1$, depending on the mean orientation being in the $x$- or $y$-directions. 
There are $2^L$ such states, for the $L \times L \times L$ lattice, say with fixed boundary conditions that enforce the specified layered order.   

When $z'=0$, the  states with different $\{\sigma\}$ are degenerate.  Now, we develop a perturbation theory  for the partition function 
$\Omega_{\{\sigma\}}( \mu,\mu, \mu')$ in powers of $ z' = \exp( \mu')$~\cite{2012-rd-pre-high,2015-nkr-jsp-high}.
\begin{equation}
\Omega_{\{\sigma\}}( \mu,\mu, \mu')=\Omega_{\{\sigma\}}( \mu,\mu, -\infty ) \exp \left[ A z' + B z'^2 + ... \right].
\end{equation}
Explicit expressions for the coefficients $A$ and $B$ can be written down in terms of expectation values of appropriate operators in the state $\{ \sigma\}$. We find that $A = \epsilon^k$, independent of $\{\sigma\}$, where $\epsilon $ is the density of holes in the problem. Hence any difference between different layer-orderings only shows up in the  $B$. 

Explicit expression for $B$ involves unperturbed probability weight of configurations that have holes that allow two $z$-rods to be put in.  Consider two z-mers that have in plane coordinates $(x,y)$ and  $(x+ \Delta_1,  y+ \Delta_2)$. The probability that we can place these two rods  breaks up into products of weights that in the $i$th layer, the sites
$(x,y,i)$ and $(x + \Delta_1,y +\Delta_2,i)$ are both not occupied by in-plane rods.  If this probability is $\alpha(\vec{r})$ or $\beta(\vec{r})$ for $x$ and $y$-orderings in the plane, with $\vec{r} = ( \Delta_1, \Delta_2)$, the term $B $ is a sum of terms of the form $\alpha^r \beta^s$, where $r$ is the number of planes with $x$-ordering that intersect both rods, and $s$ is the corresponding number of planes with $y$-ordering. By symmetry in the $x$ and $y$ directions, there will also be a term  $\alpha^s \beta^r$ for the same $\{\sigma\}$ corresponding to separation $\vec{r}'= (\Delta_2, \Delta_1)$ between the rods. But we notice that, for all $\alpha, \beta \geq 0$, and integers $r, s \geq 0$, 
\begin{equation}
\alpha^r \beta^s + \alpha^s \beta^r \leq \alpha^{r+s} + \beta^{r +s}.
\end{equation}
This implies that the second correction term, when all in-plane nematic orientations are parallel is greater than the term when they are not.  Thus, the concentration of $z$-rods induces an effective aligning interaction between nearby layers.  Note that this interaction term is proportional to the volume of the system, and would dominate over the degeneracy $2^L$  term coming from the number of different states $\{\sigma\}$.  This is an order-by-disorder mechanism, where the degeneracy between different equal-weight states $\{\sigma\}$ is lifted, once the perturbation $z'$ is introduced. 

\begin{figure}
\includegraphics[width=\columnwidth]{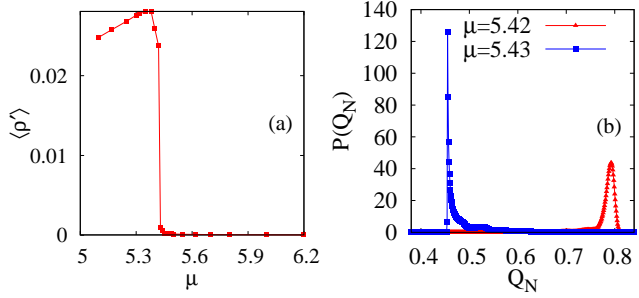}
\caption{\label{ln_stability} (a) The variation of $\langle\rho'\rangle$, the minimum of the densities of the rods of different orientations, with  $\mu$ 
for $L=112$ and $k=7$ in the vicinity of transition from nematic phase to a layered-nematic phase. 
$\rho'$ has a discontinuity as $\mu$ changes from  $\mu=5.42$ to $\mu=5.43$, representing the onset of a layered phase. The lower values of 
$\langle \rho' \rangle$ stabilizes the layered-nematic phase for finite system sizes. (b) Corresponding probability distribution $P(Q_N)$  near the vicinity of nematic-layered transition. The peak of  $P(Q_N)$ jumps as $\mu$ changes  from $\mu=5.42$ to $\mu=5.43$.}
\end{figure}

However, the excess free energy in the ordered nematic state per unit volume is only of ${\mathcal O}([\rho'/k]^2)$, where $\rho'/k$ is the number density of $z$-mers.  In our simulations, for larger values of $\mu$, $\rho'$ becomes very small, for instance in $L=112$ and $k=7$,   from Fig.~\ref{ln_stability}, we see that
$\rho'(\mu=5.42) = 0.023$ and $\rho'(\mu=5.43) = 0.0009$, just beyond onset of
the layered nematic phase, representing an order of magnitude decrease in $\rho'$ as $\mu$ is increased, so that one expects to see configurations with  non-parallel nematic order between layers with significant weight if the disordering term $L \ln 2$ is of same order as the the ordering term $L^3 \rho'^2$.  Hence, we expect that for $L > L^* \sim k/\rho'$, the ordering term will win, and nematic ordered state will dominate. However, in our simulations, for $\mu = 5.42, k=7$, $\rho'  \approx 0.023$, and so, $L^*$ is of order $300$.  For lower values of $\mu$, $\rho'$ is larger, and we do see the nematic order. In the other case of $\mu=5.43$, $L^*(\rho'=0.0009) \sim 7000$. This is much higher than the system size $L=112$, implying that
the  layered-nematic phase is favored for $\mu=5.43$.

\section{\label{sec:phasediagram}Phase diagram and critical behavior}

\begin{figure}
\includegraphics[width=0.8 \columnwidth]{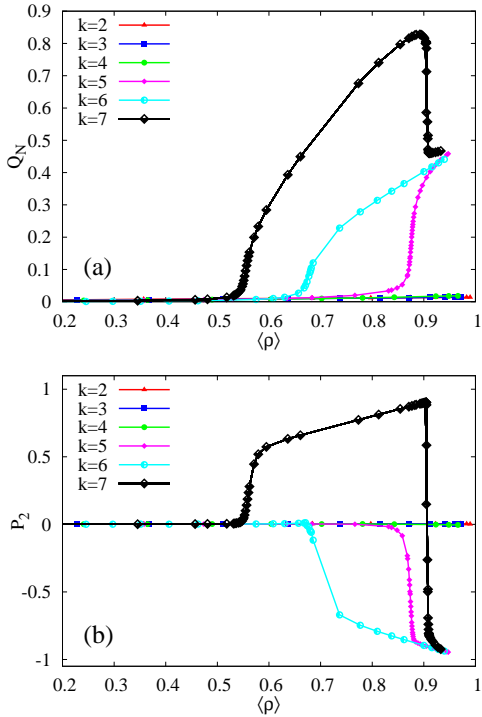}
\caption{\label{Qn} The order parameters (a) $Q_N$ [see Eq.~\eqref{eq:qn}] and  (b) $P_2$  [see Eq.~\eqref{eq:p2}] as a function
of mean density $\langle \rho \rangle$ for $k=2,\ldots,7$. The data are for systems with $L=10 k$.} 
\end{figure}

We numerically determined the order parameters 
$Q_N$ [see Eq.~\eqref{eq:qn}] and  $P_2$  [see Eq.~\eqref{eq:p2}] 
as function of $\rho$ for different $k$ as shown in Fig.~\ref{Qn}. We first determine $k_{min}$, the minimum
value of $k$ required for each of the phases to appear.

\subsection{$k_{min}$}

From Fig.~\ref{Qn}, it is evident that for $k \leq 4$, both $Q_N$ and $P_2$ are zero for all values of $\rho$. There are no phase transitions and the system is in  the disordered isotropic phase for all densities. 

For $k=5$ and $k=6$, $Q_N$ increases from 
$0$ to $0.5$ at high densities, while $P_2$ simultaneously  decreases from $0$ to $-1$. These values are indicative of the layered phase, and show that the system undergoes a single 
transition from an isotropic phase to a layered phase. 
Thus, for observing a layered phase $k\geq k^{layered}_{min}=5$. We note that there is no nematic phase when $k=5,6$.
The critical values for the isotropic-layered transition are: $\mu_c(5)\approx3.82$ and $\rho_c(5)\approx0.874$ and $\mu_c(6)\approx1.0$ and $\rho_c(6)\approx0.68$.

When $k=7$, it may be seen from Fig.~\ref{Qn} that $Q_N$ increases from zero to $\approx \rho$ and then decreases to $Q_N\approx \rho/2$. Simultaneously, $P_2$ increases to $1$ and then drops sharply to $-1$. 
These values are indicative of nematic and layered phases.
We conclude that a nematic phase exists for $k \geq k^{nematic}_{min}=7$. Numerically we find that
the layered phase may be further divided into layered-nematic and layered-disordered phases, which is presumably an artifact of the  small sizes of our 
system, as discussed in Sec.~\ref{sec:stability}.

\subsection{$k=5, 6$}

\begin{figure}
\includegraphics[width=\columnwidth]{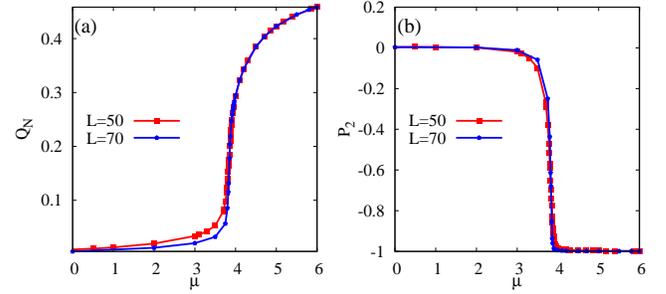}
\caption{The order parameters (a) $Q_N$ and (b) $P_2$ for $k=5$ as a function of $\mu$ for two different system sizes. The
data is very weakly dependent on the system size. }
\label{k5}
\end{figure}

Rods of length $k=5$ are the smallest to show the layered-disordered phase at high densities. 
We first show that this phase is stable and that the Monte Carlo algorithm equilibrates the system at these densities. 
To show the stability, we compare the order parameters $Q_N$ and $P_2$ for two different system size in Fig.~\ref{k5}. The data
has only a  very weak dependence on the system size, showing that the finite size effects are not important and that the layered phase is stable in the thermodynamic limit. The critical values for the transition is $\mu_c(5) \approx 3.82$ and $\rho_c(5) \approx 0.874$. 

\begin{figure}
\includegraphics[width=\columnwidth]{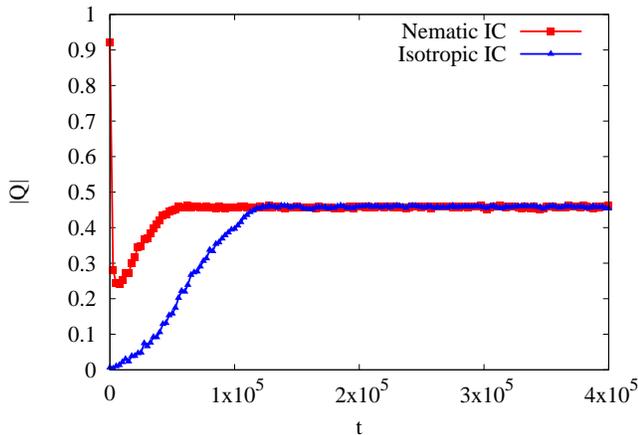}
\caption{The time evolution of $|\Qn|$ when the phase at time $t=0$ is nematic or isotropic. IC in the legends is an acronym for initial conditions.
The data are for 
$L=70$, $k=5$, and $\mu=6.0$. The system loses memory of its initial state within $10^5$ Monte Carlo steps, and
equilibrates into a layered phase characterized by $\langle |\Qn|\rangle \approx \rho/2$.}
\label{k5_stability}
\end{figure}

To check that at the high values of $\mu$ and densities $\rho$, our simulations do not suffer from slow down due to jamming problems, we observed the evolution with  two different initial conditions: one corresponding to a nematic phase and the other corresponding to an isotropic phase and check that the final state is independent of the initial conditions. The
time evolution of $|\Qn|$ is shown in Fig.~\ref{k5_stability} for both of these initial conditions. Clearly, the system loses memory of the initial conditions quite rapidly, and the order parameter reaches a value close to $0.5$, indicative of the layered phase.

\begin{figure}
\includegraphics[width=\columnwidth]{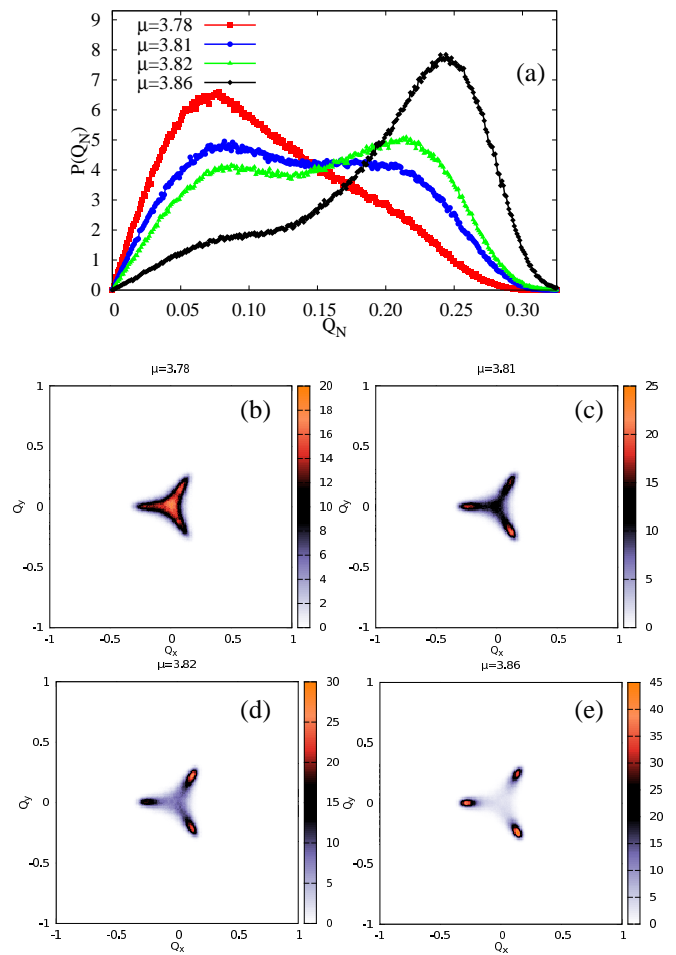}
\caption{(a) The probability distribution $P(|\Qn|)$ near the isotropic-layered transition for $L=50$ and $k=5$. (b)--(e) The two dimensional
color plots for $P(\Qn)$  for different values of $\mu$ near the isotropic layered transition for $L=50$ and $k=5$.}
\label{k5_hist}
\end{figure}

We now study of the isotropic to layered-disordered transition. There are three symmetric ordered states. By analogy to the
three state Potts model, we expect that this transition should be first order. The numerical data is consistent with a first order
transition. First, we show in Fig.~\ref{k5_hist}(a) the probability distribution of the order parameter $Q_N$ near the transition point. The distribution has two peaks for values of $\mu$ close to the transition point, one near $Q_N \approx 0$, corresponding
to the isotropic phase and the other close to $Q_N \approx 0.25$, corresponding to the layered phase. Double peaked distribution
are a signature of first order transitions and co-existence. This can be further confirmed by looking at two-dimensional density plots
of $P(\Qn)$, as shown in Fig~\ref{k5_hist}(b)--(e), where as $\mu$ is increased, the simultaneous presence of peaks at the origin and and at $\pi/3$, $\pi$ and $5 \pi/3$ can be seen in Fig~\ref{k5_hist}(c) and (d). 

\begin{figure}
\centering
\includegraphics[width=0.45\textwidth]{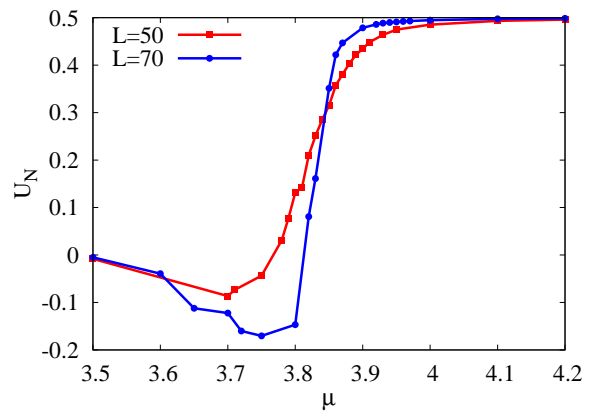}
\caption{\label{fig:binder}The variation of the Binder cumulant $U_N$ [see Eq.~\eqref{eq:binder}] with $\mu$ for two different
system sizes. The data are for $k=5$ near the isotropic-layered transition. $U_N$ becoming negative is suggestive of a first
order transition.}
\end{figure}

Further evidence of the first order nature may be obtained by studying the Binder cumulant
\be
U_N = 1-\frac{\langle|\mathbf{Q}|^4\rangle}{2\,\langle|\mathbf{Q}|^2\rangle^2}.
\label{eq:binder}
\ee
$U_N$ is zero in the isotropic phase and $1/2$ in the completely ordered phase. The variation of $U_N$ with $\mu$
is shown in Fig.~\ref{fig:binder}. The Binder cumulant becomes negative near the transition point, with its minimum decreasing
with system size. Binder cumulant becoming negative is a  strong signature of the transition being first order.

The results for $k=6$ are very similar to that for $k=5$. The system undergoes a single transition from isotropic to layered phase
with critical parameters are approximately  $\mu_c(6) \approx 1.0$ and $\rho_c(6) \approx 0.68$. We note that the critical values
are smaller than that for $k=5$.

\subsection{$k=7$}

\begin{figure}
\includegraphics[width=\columnwidth]{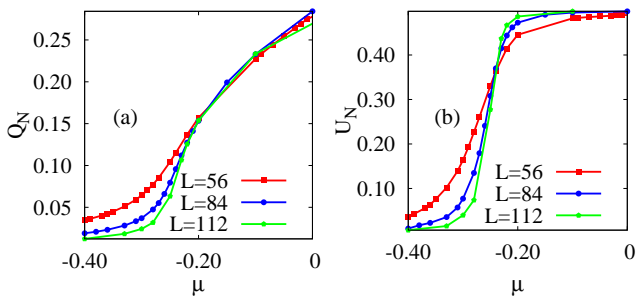}
\caption{\label{fig:IN}The variation of the (a) order parameter $Q_N$ and (b) Binder cumulant  $U_N$  with $\mu$ for three different system sizes. The curves
for the Binder cumulants cross at $\mu \approx -0.23$.   }
\end{figure}

\begin{figure}
\includegraphics[width=\columnwidth]{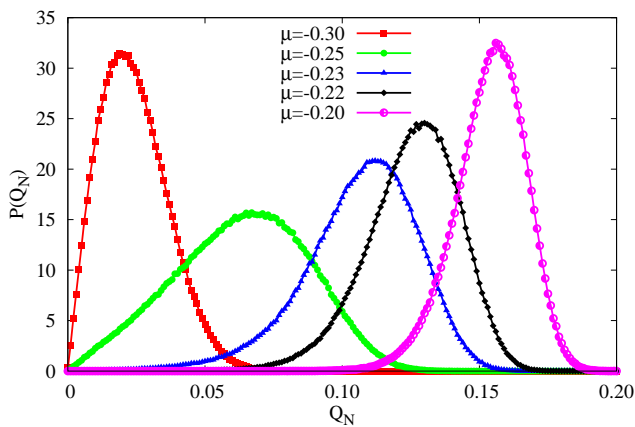}
\caption{\label{fig:INqn}Probability distribution $P(Q_N)$ for $k=7$ and $L=112$ near the I-N transition. $P(Q_N)$ is unimodal and the  peak position shifts continuously to the right with increasing $\mu$. }
\end{figure}

When $k=7$, the systems undergoes a transition from an isotropic phase to a nematic phase at low densities
and from nematic phase to a layered phase at high densities (see Fig.~\ref{Qn}). Here, we analyze the nature of the
transitions as well as the nature of the layered phase. We first discuss the isotropic-nematic transition. There are three symmetric
nematic phases corresponding to the three different orientations. By analogy with the three state Potts model, we expect that
the transition will be first order in nature. The dependence of the order parameter $Q_N$ and the Binder cumulant $U_N$ on
$\mu$ for different system sizes are shown in Fig.~\ref{fig:IN}. $Q_N$ does not show any sign of a discontinuity, nor does
the Binder cumulant become negative, both being signatures of a first order transition. Likewise, the probability distribution
for $Q_N$, shown in Fig.~\ref{fig:INqn} does not show a bimodal distribution for all values of $\mu$ near the critical point.
From the crossing of the Binder cumulants, we conclude that the critical parameters are $\mu_c \approx -0.23$ corresponding
to $\rho_c \approx 0.556$. The three state Potts model in 3-dimensions  has a very weak first order transition that is
difficult to detect in numerical simulations and we expect that the same difficulty holds for the problem of rods. Our simulations do not
five a clear evidence of the nature of this phase transition.

We now examine the transition from nematic to layered phase. For $k=7$ and $L=112$, we find a range of $\mu$ ($5.43 < \mu < 5.60$) for 
which the system finds itself in the layered-nematic phase.  We check that this phase is stable by simulating systems with
initial condition that is isotropic, nematic and layered-disordered. Also, we notice that the transition from nematic to layered nematic is accompanied by a  is a sharp decrease in $\rho'$. However, as we do not expect this to be a thermodynamic phase transition,  we did not undertake a detailed study of the layered-nematic phase.

\section{\label{sec:discussion} Summary and discussion}
  
To summarize, we studied the problem of monodispersed hard rods on a three dimensional cubic lattice using grand canonical Monte Carlo simulations and theoretical methods to obtain the phases for rods of length $k$. We showed that for $k \leq 4$, the system is in a disordered isotropic phase at all densities
$\rho$, and there are no phase transitions. For $k=5, 6$, the system undergoes a single transition into a high density layered-disordered phase, where the system breaks up into two dimensional layers, but disordered within a layer. For $k=7$, we find that as density is increased, the system  makes a transition into a nematic phase. Further increase of density results in  a layered-disordered phase.  We also observe a layered-nematic phase between the nematic and layered-disordered phases, which we have argued is a finite-size effect.

\begin{figure}
\includegraphics[width=\columnwidth]{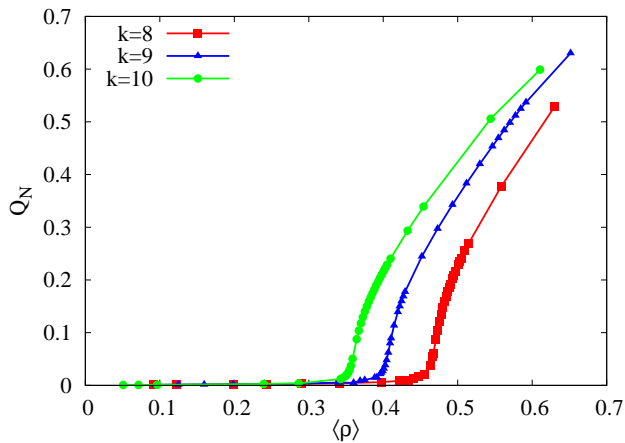}
\caption{\label{fig:largerk} The behavior of the order parameters $Q_N$  near the isotropic-nematic transition
for $k=8,9,10$. The data are for $L=10 k$. }
\end{figure}

For values of $k>7$, we expect that the phase diagram remains qualitatively the same as that for $k=7$. We expect 
the critical density for the isotropic-nematic transition to decrease with increasing $k$, as is confirmed by Monte Carlo
simulations of systems with $k=8,9,10$. As seen from Fig.~\ref{fig:largerk}, $\rho_c$ decreases from 0.556 for $k=7$ to 0.364 for $k=10$. For large $k$, we expect that $\rho_c \sim k^{-2}$, as can be seen by assuming that near the transition point, number of rods per $k \times k \times k$ cube  should be of 
of order 1. The nematic-layered transition is essentially a  two dimensional transition, as different layers are nearly independent. Thus, we expect that the critical density for this transition varies as $1-a/k^2$ for large $k$ as in two dimensions~\cite{2007-gd-epl-on}.

These arguments are  easily extended  to  higher dimensions. For large enough $k$, we will expect  a isotropic-nematic
transition at a critical density that scales as $k^{-(d-1)}$. A nematic phase may be thought of a union of parallel lines, with  hard core constraint along a line, and the problem 
becomes essentially  one  dimensional, with weak correlations between different lines.
In the high-density phase, we expect that the system will  break to {\it  two-dimensional} layers, with only weak interaction between different layers. This
critical density will vary as $1-a/k^2$ for large $k$. Preliminary simulations in four dimensions are consistent with
the above observations.

From the results of the paper, it is clear that for $k\geq 5$, the fully packed phase  shows spontaneous symmetry breaking by selecting the  layering plane. It is thus qualitatively different from the $k \leq 4$. 
Extending the problem of rods to cuboids would result in a much richer phase diagram, as expected from the corresponding case of hard rectangles in two dimensions.  However, simulations of such systems is a  challenging task.

\begin{acknowledgments}
The simulations were carried out in single node cluster machines Annapurna(2 x Intel Xeon Processor X5570 2.93 GHz) and Nandadevi(2 x Intel Xeon E5-2667  3.3 GHz) using OpenMP C language. We thank Joyjit Kundu and Kedar Damle  for helpful discussions. 
DD's work is partially supported by the grant DST-SR-S2/JCB-24/2005 of the Government of India.
\end{acknowledgments}

\end{document}